# Machine-learned potential for amorphous Indium-Tin-Oxide alloys

Shuaiyang Guo[1], Yuan Wang[1] and Wei Zhang[1,*]

[1]Center for Alloy Innovation and Design (CAID), State Key Laboratory for Mechanical Behavior of Materials, Xi'an Jiaotong University, Xi'an, China

E-mail: wzhang0@mail.xjtu.edu.cn

**Abstract:**

Machine-learned potential-driven molecular dynamics (MLMD) simulations are of great value in guiding the design and optimization of memory devices. Amorphous indium-tin-oxide (ITO) is widely used as transparent conducting oxide for flat-panel display and solar cell applications, and also as a capping layer in phase-change-materials-based reconfigurable color display devices. However, atomistic simulations of ITO using ab initio molecular dynamics (AIMD) are limited to systems of a few hundred atoms due to expensive computational costs, which prevents the device-scale modelling of real-world applications. In this work, we develop a machine-learned potential for ITO and its parent phase $In_2O_3$ based on the Gaussian approximation potential (GAP) framework. We generate a comprehensive training dataset using an iterative training protocol. Our MLMD simulations of crystalline, liquid and melt-quenched amorphous ITO models are in great agreement with the AIMD reference. In particular, the ML potential well captures the minority atomic interaction, such as Sn–Sn bonds, which have poor statistics in small-scale AIMD simulations. We demonstrate that the MLMD simulations are 3–4 orders of magnitude faster than AIMD. The training dataset and the fitted GAP potentials for ITO and $In_2O_3$ are openly accessible.

**Keywords:** indium-tin-oxide, machine-learned potential, Gaussian approximation potential, amorphous phase, MLMD

# 1. Introduction

Transparent conducting oxides (TCOs) are a class of materials characterized by a wide optical bandgap, high electrical conductivity, high optical transmittance, and compatibility with low-temperature fabrication processes [1-5]. These properties enable their widespread use in thin-film transistors [6-11] and transparent electrodes [12, 13]. Among them, indium-tin oxide (ITO) with an In-rich composition, e.g., $In_2O_3$:$SnO_2$ = 90:10 wt.%, exhibits an optical bandgap of ~3.8 eV, an electrical conductivity of ~$10^4$ S·cm$^{-1}$, and a high transmittance above 90% in the visible spectrum [14-18]. For phase-change material (PCM)-based applications [19-26], ITO is commonly employed as electrodes or capping layers for reconfigurable optical devices. By applying electrical or optical pulses, the PCM layer in the ITO/PCM/ITO sandwich device can be switched between the amorphous and crystalline phases for reconfigurable color display. The base color of the sandwich device can be effective tuned across the visible-light range by varying the thickness of the bottom ITO layer [27-33]. Density functional theory (DFT) and DFT-based *ab initio* molecular dynamics (AIMD) simulations were reported to gain atomic-scale understanding of the structural and optical properties for both crystalline and amorphous ITO [34-39]. However, AIMD simulations are typically limited to systems of a few hundred atoms, which prevents the direct device-scale modelling of ITO-based devices.

Machine-learned (ML) potential can substantially improve computational efficiency while retaining the *ab initio* computing accuracy [40-46]. An ML potential is a mathematical representation of the potential energy surface (PES) of the system, mapping the energy of a given system to its local atomic structure. In the past years, various ML potentials have been successfully developed for the key functional materials used in electronic memory and display devices. In addition to the potentials developed for binary oxides such as $SiO_2$ [47-51], $HfO_2$ [52-54], $CeO_2$ [55], $Ga_2O_3$ [56-58] and $In_2O_3$ [59], large-scale MLMD simulations have been performed for chemically and structurally complex PCMs including pure Sb [60, 61], GeTe [62-65], Sb–Te alloys [66, 67], $Ge_2Sb_2Te_5$ [68, 69], $Ge_3Sb_6Te_5$ [70] and all the Ge–Sb–Te compositions along the GeTe–$Sb_2Te_3$ pseudo-binary line [71, 72], enabling atomistic simulations at real-world device scales [71-76]. Our previous work demonstrated full-cycle device-scale simulations including the RESET (amorphization) [71, 72] and the SET (crystallization) operation [72] of a 20 × 20 × 40 nm$^3$ Ge–Sb–Te model (containing over 530,000 atoms), corresponding to the size of a single memory cell in the cross-point memory product [77].

We note that ITO serves as a crucial component in a wide range of electronic and photonic devices; however, to date, no general-purpose ML potential has been developed, which is capable of simulating the crystalline, amorphous, and liquid phases of ITO. In this work, we perform thorough DFT and AIMD calculations to investigate the structural, electronic and optical properties of amorphous and crystalline ITO. The DFT-calculated refractive index and extinction coefficient yield close comparisons with recent spectroscopic ellipsometry data [78]. We develop an efficient ML potential with *ab initio* accuracy for ITO and $In_2O_3$ via the iterative training procedure under the Gaussian approximation potential (GAP) scheme [79]. We make thorough structural analyses of ordered and disordered ITO structures and confirm that even the atomic interactions between minority atoms can be well reproduced by the ML potential. We also show that the GAP model can be used to simulate other commonly used ITO compositions, despite that these structures are not included in the training dataset. Finally, we study the structural features of amorphous ITO from a few hundreds of atoms to 10,000 atoms, and test the computing efficiency of the GAP model up to 700,000 atoms.

## 2. Methods

We used the Vienna Ab-initio Simulation Package (VASP) [80] for DFT and AIMD (Born-Oppenheimer MD) simulations. The projector augmented wave (PAW) pseudopotentials [81] and the Perdew-Burke-Ernzerhof (PBE) functional [82] were employed. If not specified, ITO is abbreviated for the composition with $In_2O_3:SnO_2$ = 90:10 wt.%. For structural relaxation of crystalline $In_2O_3$ and ITO supercell models, the *k*-point meshes were set as 2×2×2 and 1×2×2, respectively. The AIMD amorphous models were generated using the canonical ensemble (NVT) and the Nosé-Hoover thermostat [83, 84] with a time step of 2 fs. The Brillouin zone of amorphous models was sampled at the Gamma point only. For electronic structure and optical calculations, the *k*-point meshes were enlarged by three times for all the crystalline and amorphous models. The Grimme's D3 method was used to account for van der Waals corrections [85]. The energy cutoff for plane waves was set as 450 eV for the DFT and AIMD calculations. The ML potential was fitted using GAP framework [79] as implemented in the QUIP code (https://github.com/libAtoms/QUIP). The per-structure energies and per-atom forces of all training configurations in the reference dataset were computed via self-consistent DFT calculations. An energy cutoff of 550 eV was used, and the energy tolerance was set as $1×10^{-7}$ eV. The GAP-based MLMD simulations were performed using LAMMPS code [86] with an interface to the QUIP code. The amorphous models were generated following the same melt-quench scheme as that in the AIMD simulations. The MLMD simulations were carried out in the NVT ensemble using the Langevin thermostat [87] with a timestep of 2 fs.

## 3. Results and Discussion

Figure 1(a) shows the DFT-relaxed atomic structure of crystalline (*c*-) $In_2O_3$, with a cubic bixbyite structure ($Ia\bar{3}$ space group). The unit cell contains 32 In atoms, 48 O atoms and 16 vacant sites. Each In atom is coordinated with six neighboring O atoms, and the In–O bond length ranges from 2.12 Å to 2.22 Å. We constructed a 2×1×1 supercell of cubic $In_2O_3$ to incorporate more Sn atoms for improved statistics, as shown in figure 1(b). In total, 6 In atoms were replaced with Sn atoms, and the In:Sn ratio of the *c*-ITO model is approximately 9.77:1. Upon structural relaxation, the volume of the *c*-ITO model was slightly expanded due to the presence of Sn atoms. Figure 1(c) and 1(d) show the amorphous (*a*-) models of $In_2O_3$ (72 In atoms, 108 O atoms) and ITO (70 In atoms, 7 Sn atoms, 119 O atoms), which were obtained via melt-quench AIMD simulations. The initial $In_2O_3$ and ITO structures were first randomized at 3000 K for 10 ps, cooled down to ~2300 K (the melting point of $In_2O_3$) at a rate of 100 K/ps, and kept at this temperature for 30 ps. A subsequent rapid quench to 300 K at a rate of 200 K/ps was performed, followed by another 30 ps equilibrium calculation at 300 K for data collection of the amorphous structures. Prior to the electronic structure and optical properties calculations, the amorphous models were cooled down to 0 K and fully relaxed. As shown in figure 1(d), both In and Sn atoms in *a*-ITO form distorted octahedral motifs.

Figure 1(e) shows the density of states (DOS) of *c*- and *a*-$In_2O_3$, which are in good agreement with previously reported theoretical results [88-90]. Regarding ITO, the presence of Sn impurities induces more mid-gap states, and the overall bandgap of *a*-ITO is smaller than that of *c*-ITO. Figure 1(g) shows the refractive index (*n*) and extinction coefficient (*k*) of *c*- and *a*-$In_2O_3$ computed in the range between 200 nm and 2200 nm. The optical profiles of the two phases are quite similar at smaller wavelengths. A major difference is observed above 1440 nm, and in particular, the *k* values of *c*-$In_2O_3$ are still close to zero, but those of *a*-$In_2O_3$ are gradually increased to ~0.6 at 2200 nm. Regarding ITO, the *c*-phase shows a notable reduction in *n* above 400 nm and a clear increase in *k* above 1000 nm, as compared to the *a*-phase (figure 1(h)). These DFT-computed optical features of the *c*-ITO and *a*-ITO models are consistent with experimental observations of ITO thin films [78].

Next, we develop an ML potential for ITO to enable large-scale MLMD simulations. We note that constructing a reliable training dataset is the first, and the most critical step for fitting an ML potential [40]. Rather than relying on automated *de novo* exploration strategies [91] or active-learning algorithms [52] that search the PES in a purely data-driven way, we manually constructed our training dataset based on specific domain knowledge of ITO and its amorphous phase. We note that this expert-guided curation approach remains as a standard and essential practice in the development of ML potentials for practical domain applications [92]. We combined the Smooth Overlap of Atomic Positions (SOAP) descriptor [93] with the GAP framework [79], a combination proven to be data-efficient and robust for describing complex PES [91, 94].

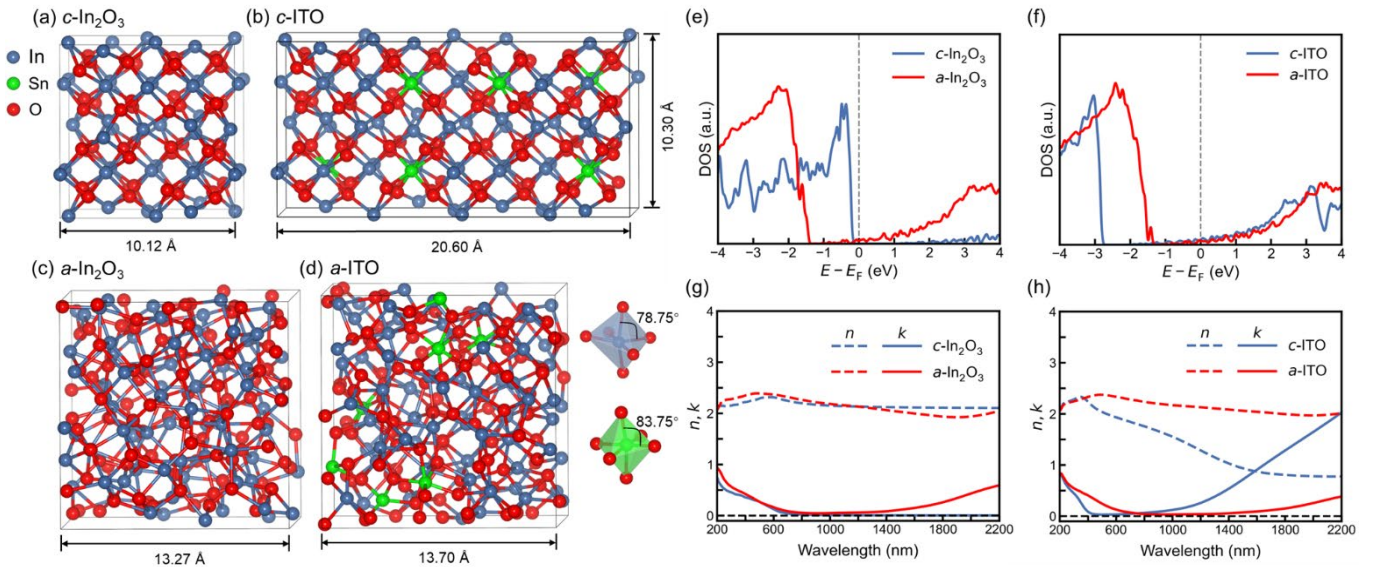

**Figure 1.** DFT-relaxed crystalline structure of (a) $In_2O_3$ and (b) ITO. Atomic structures of melt-quenched amorphous models of (c) $In_2O_3$ and (d) ITO. The typical local motifs of In and Sn atoms in amorphous ITO are displayed. The DFT-calculated density of states for (e) $In_2O_3$ and (f) ITO. The DFT-calculated refractive index *n* and extinction coefficient *k* for (g) $In_2O_3$ and (h) ITO.

Figure 2(a) shows the fitting process of the ITO GAP ML potential. The reference dataset was progressively expanded to improve the ability of the potential to describe complex structures. The initial dataset (iter0) includes three isolated atoms (In, Sn and O), dimer, crystalline, and AIMD-generated melt-quenched amorphous configurations. Regarding crystalline structures, we considered the unit cell models of In, Sn, O, $In_2O_3$ and $SnO_2$, and supercell models of ITO. We also included multiple crystalline configurations with additional structural distortions and some hypothetical phases obtained from Materials Project database [95]. Many disordered structural snapshots during the high-temperature melting and the rapid-quenching processes were included in the dataset, in addition to the amorphous structures obtained at 300 K. In total, 666 training structures were used to fit an initial GAP model (iter0).

Two iterative training processes were performed, in which the previous version of the GAP model was used to perform melt-quench simulations and generate new configurations for the subsequent fitting. In the first iteration, we added 40 new disordered structures of $In_2O_3$, $SnO_2$ and ITO sampled at their respective experimental densities [96-98], and fitted the iter1 potential. We then repeated this process to expand the compositional and configurational space in the second iteration: we generated structures for three distinct amorphous oxide compositions (viz. $In_2O_3$, $SnO_2$, and ITO) with densities varying by ±5% relative to the experimental values, leading to another 90 training configurations. In total, the final reference dataset comprises 796 simulation cells containing 65,218 atoms (table 1), leading to complex structural and chemical space (figure 2(b)). Figure 2(c) shows the numerical validation of the final ITO-GAP model: the GAP-predicted energies and forces exhibit excellent agreement with the DFT references, indicating a great numerical accuracy.

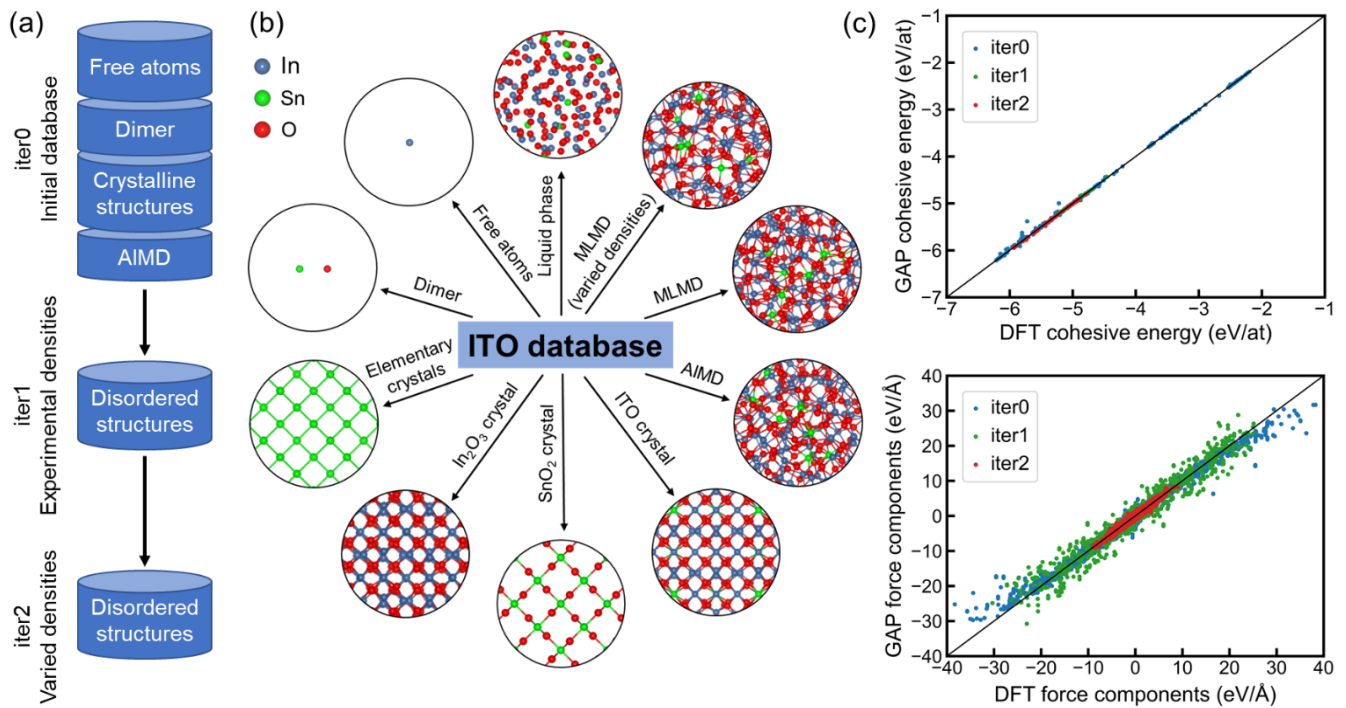

**Figure 2.** The GAP fitting process. (a) Construction of the reference dataset via iterative training. (b) Representative structures of the dataset. (c) ML-predicted versus DFT-calculated cohesive energy and force.

**Table 1.** Details of simulation cells, number of atoms and sparse points employed for the GAP fitting.

|  |  | Data size ||  Sparse points |
|---|---|---|---|---|
|  |  | Cells | Atoms |  |
| Free atoms | In / Sn / O | 3 | 3 | 1 |
| Dimer data | In-In / In-Sn / In-O / Sn-Sn / Sn-O / O-O | 211 | 422 | 90 |
| iter0 | Crystalline structures | 361 | 23813 | 1500 |
|  | AIMD structures, from melt-quench simulation | 91 | 16860 | 1500 |
| iter1 | Disordered structures (experimental densities) | 40 | 7440 | 600 |
| iter2 | Disordered structures (variable densities) | 90 | 16680 | 1300 |
| **Total** |  | **796** | **65218** | **4991** |

We then performed thorough structural validations for our ITO-GAP potential. To ensure a direct comparison with AIMD reference data, we used identical simulation cell dimensions and the same number of atoms. Trajectories were collected for crystalline and amorphous models at 300 K, and for liquid models at 2300 K. The crystalline systems comprised 160 atoms, while the amorphous and liquid systems contained 196 atoms. Figure 3 presents the radial distribution functions (RDF), angular distribution functions (ADF), and coordination number (CN) distributions computed from AIMD and MLMD trajectories. For liquid and amorphous ITO, the results were averaged over 5 different AIMD / MLMD runs with independent thermal histories. For crystalline ITO, the same initial configuration was used for the AIMD and MLMD. The cutoff value for the In–In, In–Sn, In–O, Sn–Sn, Sn–O and O–O contact was set to be 3.08, 3.01, 2.70, 2.96, 2.67 and 2.42 Å, respectively. We show that our GAP model accurately reproduced the structural features of the AIMD references. We also performed the structural analyses for crystalline, liquid and amorphous $In_2O_3$, which yielded similarly excellent agreement (figure S1).

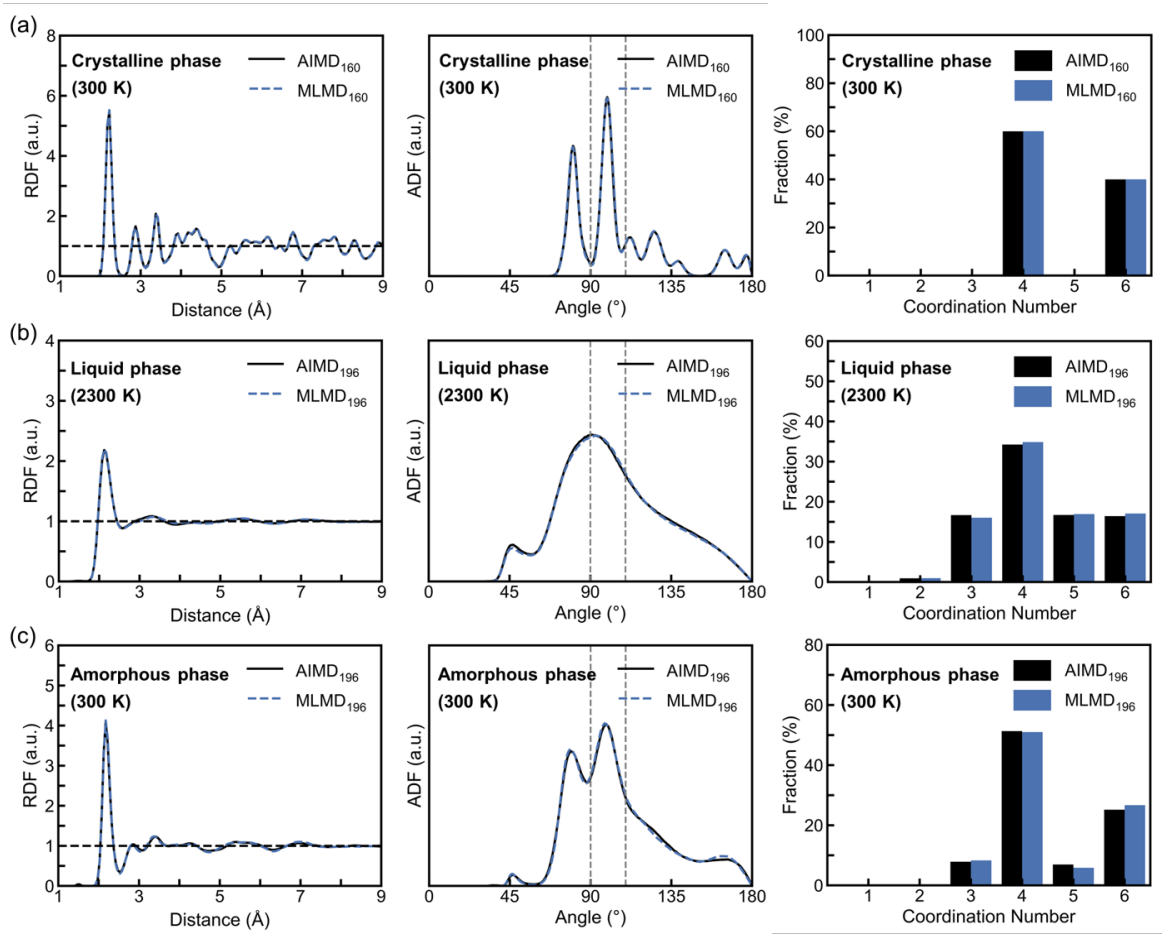

**Figure 3.** The radial distribution function, angular distribution function and distribution of coordination number calculated via AIMD and MLMD for (a) crystalline (300 K), (b) liquid (2300 K) and (c) amorphous (300 K) ITO models.

We note that accurately capturing the interactions of minority atomic species is a challenge for AIMD due to limited simulation size. In small simulation cells (up to several hundred atoms), minority species are present in such low numbers that their spatial distribution is often dominated by statistical noise, leading to finite-size artifacts that obscure the description of local structural environments. We computed partial RDFs for $a$-ITO using only a single set of AIMD / MLMD trajectory. As shown in figure 4, the heteropolar In–O and Sn–O bonds are the dominant bonding pairs in $a$-ITO with a major peak appearing at 2.18 Å and 2.08 Å, respectively. Regarding the homopolar In–In and O–O bonds, their primary peak appears at 3.43 and 2.83 Å, respectively. Overall, the MLMD and AIMD results are nearly identical for these major pair interactions. However, notable discrepancies are observed in the partial RDFs for Sn–In and Sn–Sn pairs. These differences stem from the poor statistics inherent to the small-scale AIMD reference: with only 7 Sn atoms in the 196-atom $a$-ITO model, the spatial distribution of Sn atoms (highlighted in figure 5a) varies largely between independent melt-quench runs, resulting in large variations in the partial RDF of Sn–In and Sn–Sn.

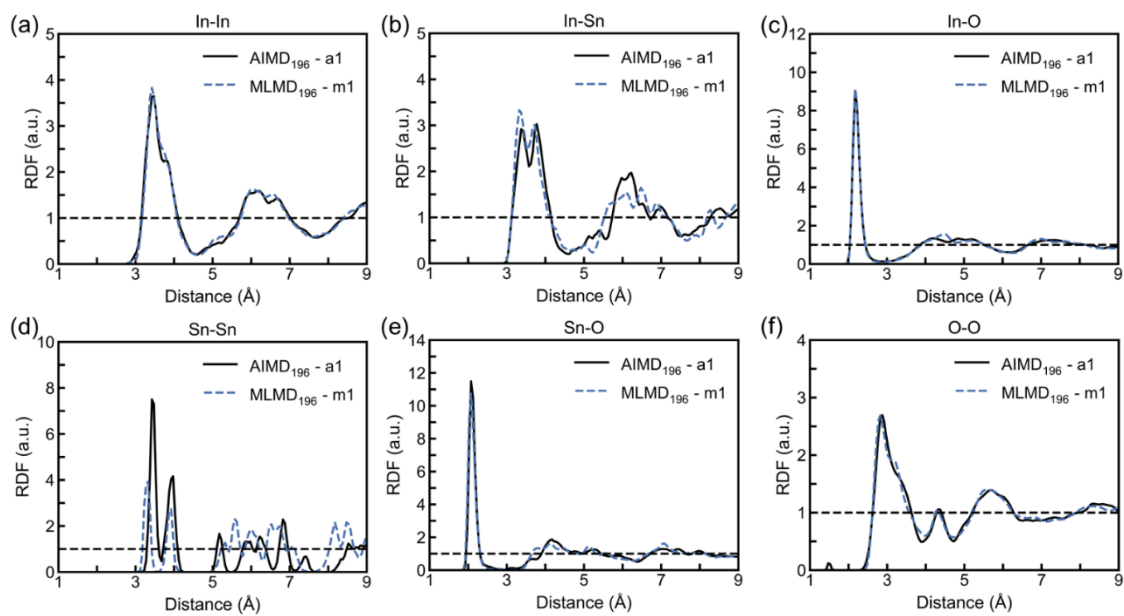

**Figure 4.** The partial RDFs of amorphous ITO models for (a) In–In, (b) In–Sn, (c) In–O, (d) Sn–Sn, (e) Sn–O and (f) O–O contacts. The AIMD$_{196}$-a1 and MLMD$_{196}$-m1 were generated through independent melt-quenched MD runs, and only this set of models was used for data collection.

To further investigate the finite-size effect, we generated four additional *a*-ITO models using AIMD and four using MLMD via independent melt-quench simulations. As shown in figure 5(b) and 5(c), the Sn–Sn RDFs across these ten independent runs exhibit large variations, indicating that the discrepancy is driven by the limited number of Sn atoms in the small simulation cells rather than inaccuracies of the potential. We performed separate AIMD and MLMD calculations at 300 K for 20 ps, starting from the exact same initial amorphous structure. In this direct comparison, the Sn–Sn RDFs show excellent agreement between AIMD and GAP-MD (figure 5(d)), demonstrating that our GAP potential has achieved *ab initio* accuracy. Finally, we increased the model size of *a*-ITO to include more Sn atoms. As shown in figure 5(e), the Sn–Sn RDF converges to a consistent profile for models containing 1,960 and 7,056 atoms, indicating that the finite-size effect is eliminated and the sampling of minority Sn–Sn interactions becomes statistically sufficient at these large scales.

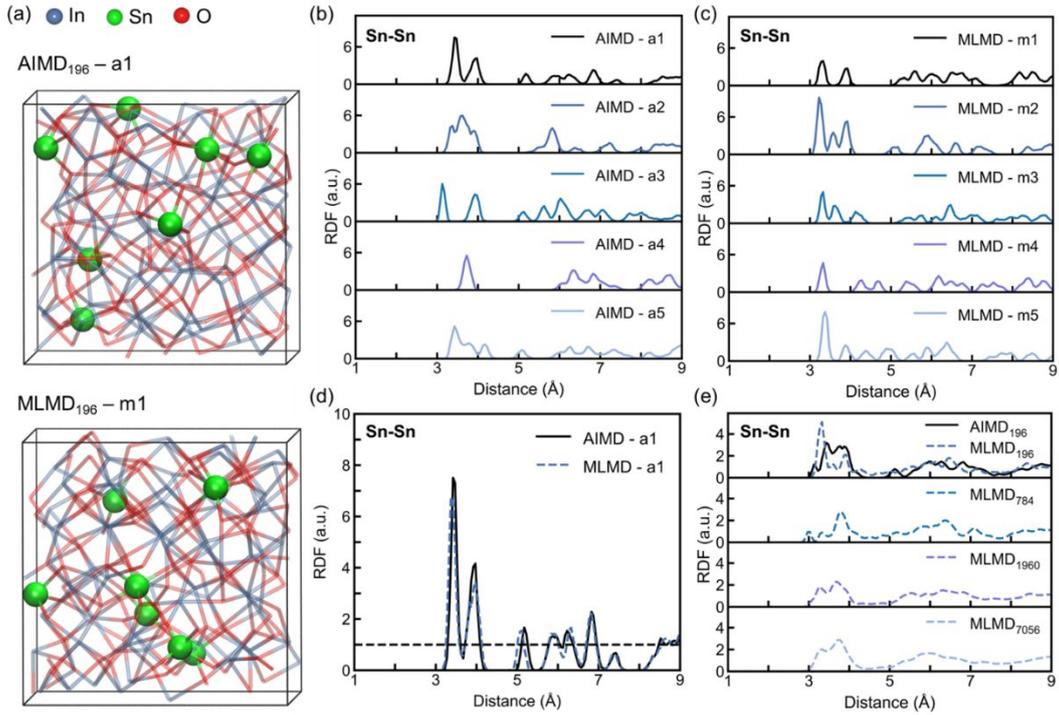

**Figure 5.** Statistical investigations of the Sn–Sn contacts in *a*-ITO models. (a) The snapshots of AIMD$_{196}$-a1 and MLMD$_{196}$-m1 models with Sn atoms being highlighted. The RDFs of Sn–Sn contacts for 5 independent *a*-ITO models generated using (b) AIMD and (c) MLMD, respectively. (d) The comparison of Sn–Sn RDF calculated using AIMD and MLMD with the same initial amorphous structure. (e) The Sn–Sn RDF of *a*-ITO models with different sizes.

Although only one representative ITO composition (i.e., In$_2$O$_3$:SnO$_2$ = 90:10 wt.%) was included in the reference dataset, the resulting GAP model demonstrates considerable chemical transferability to other experimentally relevant compositions, such as In$_2$O$_3$:SnO$_2$ = 95:5 wt.%, 85:15 wt.%, and 80:20 wt.%. Figure 6 compares various structural properties computed via MLMD and AIMD for these three amorphous alloys. The results are in good agreement across all metrics. Since our GAP potential accurately describes both pure In$_2$O$_3$ and the 90:10 wt.% ITO alloy, compositions falling between these two points are well interpolated. Regarding higher Sn concentrations, we only validated the GAP model up to a composition of 80:20 wt.%. While the current model is not intended to span the entire pseudo-binary line, we note that the flexibility of the GAP framework allows for straightforward extension to these compositions by adding additional DFT-labeled configurations into the training dataset as needed.

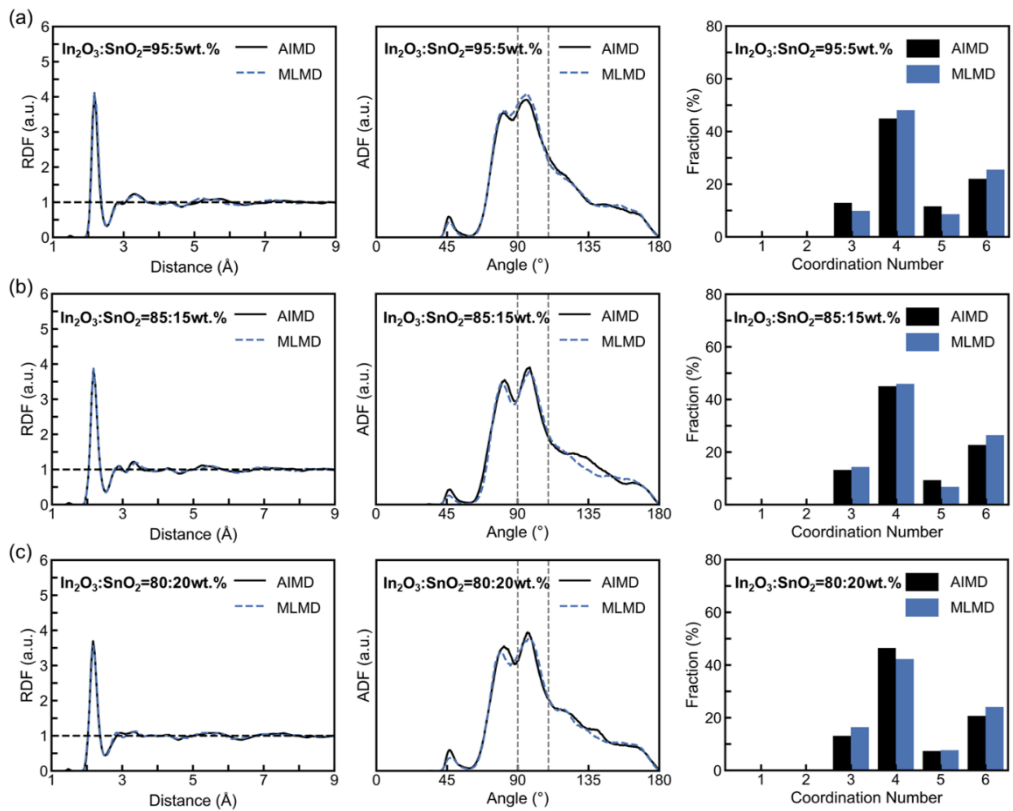

**Figure 6.** Structural analyses of with slightly different ITO compositions, namely (a) $In_2O_3:SnO_2$ = 95:5 wt.%, (b) $In_2O_3:SnO_2$ = 85:15 wt.% and (c) $In_2O_3:SnO_2$ = 80:20 wt.%, respectively. The structural data were collected from the MD trajectory at 300 K of a single amorphous model.

Finally, to quantify the computational efficiency of MLMD, we performed scaling tests comparing AIMD and MLMD using the Computing Center in Xi'an platform. Each computing node is equipped with Hygon 7285H processors (64 CPU cores) and 256 GB of memory. Figure 7(a) displays the performance on 8 nodes (512 cores in total) across varying model sizes. We observe that MLMD exhibits approximately linear scaling for large-scale models (>10,000 atoms), whereas smaller systems (<1,000 atoms) show a deviation from the ideal trend due to parallelization overhead. Notably, the computational advantage of MLMD over AIMD becomes increasingly pronounced as the system size grows. While the AIMD efficiency decreases from 14,940 MD steps per day for a 100-atom system to just 24 steps per day for a 2,000-atom system, MLMD achieves 218,796 steps per day for the same 2,000-atom model. This corresponds to an acceleration of approximately 9,000 times. With only 8 nodes, we successfully scaled the MLMD simulations to systems containing approximately 700,000 atoms. Similar speed-up trends were observed in the tests performed on 4 nodes (figure S2). Figure 7(b) presents amorphous ITO models of increasing size (196, 1,000, and 10,000 atoms); the consistency of their total RDFs confirms that the structural description remains robust across different length scales.

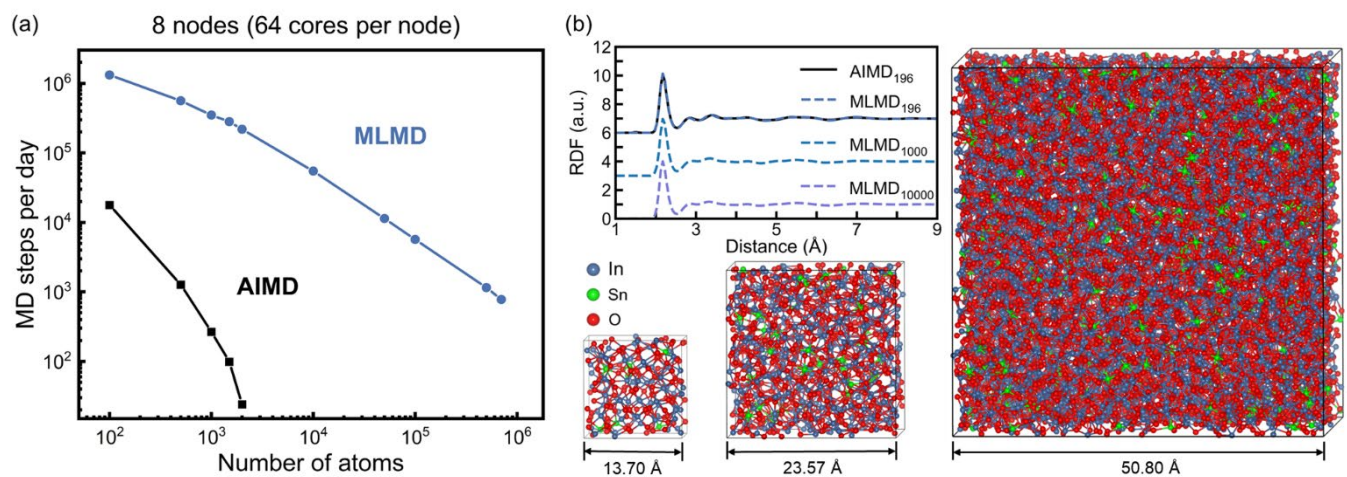

**Figure 7.** The scaling tests. (a) The MD steps per day as a function of system size running on 8 computing nodes with 64 CPU cores per node. The scaling analysis stopped at approximately 700,000 atoms due to hardware memory limitations on the compute nodes. (b) Atomistic snapshots of amorphous ITO models containing 196, 1,000, and 10,000 atoms (from left to right), shown together with their corresponding total radial distribution functions.

## 4. Conclusion

In this work, we combined DFT and MLMD simulations to comprehensively investigate the structural, electronic, and optical properties of ITO, an important material for advanced memory and optoelectronic applications. Our electronic structure calculations reveal that Sn substitution shifts the Fermi level in the crystalline phase, generating a distinct optical contrast with the amorphous phase in the near-infrared region, which is a feature absent in pure $In_2O_3$. To bridge the efficiency gap between classic AIMD and device-scale modelling, we developed a machine-learned potential for the In–Sn–O system using the GAP framework. This GAP model achieves *ab initio* accuracy with substantially higher computational efficiency, enabling MLMD simulations of systems up to hundreds of thousands of atoms. We found that these large-scale MLMD simulations eliminated the statistical noise inherent to minority Sn–Sn interactions in smaller models, indicating that standard AIMD cell sizes are insufficient for capturing the local structure of low-concentration dopants. This transferable and extensible potential now provides a robust platform for investigating the atomic-scale operation and failure mechanisms of ITO-based devices.

## Data availability

The potential files, fitting dataset and the amorphous snapshots shown in figure 7 in this work are will be available for free access via Zenodo upon journal publication.

## Conflicts of interest

The authors declare no conflict of interest.

## Acknowledgements

The Computing Center in Xi'an is acknowledged for providing computational resources. The authors acknowledge the support of XJTU for their work at CAID.